\documentclass[twocolumn]{aastex701}

\usepackage[utf8]{inputenc}
\usepackage{graphicx}
\usepackage{bm}
\usepackage{bm}
\usepackage{makecell}
\graphicspath{./figs/}
\usepackage{graphicx}	
\usepackage{amsmath}	
\usepackage{amssymb}
\usepackage{xcolor}
\usepackage{enumitem}
\usepackage{tcolorbox}
\tcbuselibrary{breakable}
\usepackage[capitalise]{cleveref}
\usepackage{subcaption}

\def\rmin{r_{\rm min}}
\def\rmax{r_{\rm max}}

\def\Ro{\mbox{\rm Ro}}

\def\Rs{R_{\odot}}

\crefname{section}{section}{sections}
\Crefname{section}{Section}{Sections}

\begin{document}

\title{Solar differential rotation driven by baroclinic forcing}

\author[orcid=0000-0002-3875-3345]{Menicucci, L.S.,}
\affiliation{Physics Department, Universidade Federal de Minas Gerais, Av. Antonio Carlos, 6627, Belo Horizonte, MG 31270-901, Brazil}
\email[show]{lucamenimeni@gmail.com}  

\author[orcid=0000-0002-2671-8796]{Guerrero, G.} 
\affiliation{Physics Department, Universidade Federal de Minas Gerais, Av. Antonio Carlos, 6627, Belo Horizonte, MG 31270-901, Brazil}
\email{guerrero@fisica.ufmg.br}

\author[orcid=0000-0002-2227-0488]{Dikpati, M.} 
\affiliation{High Altitude Observatory, NSF  National Center for Atmospheric Research, Boulder, Colorado, USA}
\email{dikpati@ucar.edu}

\author[orcid=0000-0003-4892-3835]{Hester, R.} 
\affiliation{Department of Physics and Astronomy, George Mason University, 4400 University Drive, Fairfax, VA 22030, USA}
\email{}

\author[orcid=0000-0003-0951-2486]{Zhang, J.} 
\affiliation{Department of Physics and Astronomy, George Mason University, 4400 University Drive, Fairfax, VA 22030, USA}
\email{}

\author[orcid=0000-0001-7077-3285]{Smolarkiewicz, P.K.} 
\affiliation{National Center for Atmospheric Research, Boulder, Colorado, USA}
\email{smolar@ucar.edu}

\begin{abstract}
	A combination of recent observations and numerical simulations has called into question whether Sun's interior is characterized by strong turbulent convection, a problem known as convective conundrum. In light of a possible absence of vigorous convective motions, we examine whether the Sun’s differential rotation, previously assumed to be tightly coupled to Reynolds stresses, may instead be 
    sustained by the presence of a background latitudinal entropy gradient in thermal wind balance. By performing global hydrodynamical simulations of a rotating spherical shell representing the bulk of the solar convection zone, we demonstrate that solar-like differential rotation can be generated under a variety of thermal stratifications. This work proposes an alternative scenario for the origin of solar differential rotation that accommodates the previously reported discrepancies.
\end{abstract}


\section{Introduction}
\label{sec:intro}

Solar plasma in the polar regions takes roughly 32 earth days to perform one rotation, at the same radius but at the equator it takes about 27 earth days, a phenomenon known as differential rotation. Over the past decades, helioseismology has revealed that the profile of angular velocity from pole to equator is monotonic and persists at each spherical shell down to the tachocline, at $r \sim 0.71R_\odot$, \citep{larson_global-mode_2018}, where an abrupt transition occurs from a latitudinally varying profile to an almost perfect solid body rotation. These large scale features are expected to play an important role in the regeneration of toroidal magnetic field during the solar cycle \citep{charbonneau_dynamo_2020}, yet they remain remarkably steady on the cycle's timescale. On the other hand, the same plasma is responsible for transporting the excess heat generated by nuclear reactions up to the surface. Because these heat driven currents occur asymmetrically to the rotation axis, they can transport angular momentum and contribute to the occurrence of the observed differential rotation.

Ever since their popularization, numerical simulations have tried to understand the maintenance of a steady rotation profile as a consequence of vigorous turbulent convection \citep{gilman_nonlinear_1977}. The long-standing effort has provided many insights regarding the convection in rotating spherical shells under the variation of several parameters \citep[e.g.,][]{brun_turbulent_2002,miesch_solar_2006,featherstone_meridional_2015,omara_velocity_2016,hindman_morphological_2020,guerrero_implicit_2022,camisassa_solar-like_2022}. Among these works, a consensus has been established that, in order to obtain a solar-like rotation profile, one has to adjust the simulations parameters to obtain a low Rossby number, $\Ro \ll 1$, corresponding to a rotationally constrained convection. By the different ways in which this can be accomplished, one ends up outside the Sun's expected parameter regime \citep[refer to][for a comprehensive discussion]{kapyla_simulations_2023}. Additionally, the origin of the tilted isocontour of rotation remains unsolved.

Meanwhile, helioseismic measurements \citep{hanasoge_anomalously_2012,birch_solar_2024,hanson_supergranular-scale_2024,stefan_time-dependence_2026} have reinforced the discrepancy between the power of sub-photospheric convection and the one predicted by the Mixing Length Theory (MLT), the orthodox theory for modeling the convective transport of energy in stellar structure models \citep{bohm-vitense_uber_1958,kippenhahn_stellar_2012,joyce_review_2023}. In the view of MLT, a convective flux $F_c$ takes place when opacity is high enough as to produce a Schwarzschild unstable stratification, and its value is proportional to the local pressure scale height. The proportionality constant is a free parameter characterizing the typical vertical flow's scale, which does not match the observed spectra peak. Also, MLT prescribes a symmetry between upflows and downflows, a property clearly not satisfied by photospheric flows \citep{nordlund_solar_2009}.

The conjuncture of problems regarding the inability of simulations to reconcile a vigorous convection, as predicted by MLT, with both the subsurface convective spectra and the differential rotation observations is referred to as the convective conundrum \citep{omara_velocity_2016}. An alternative framework which accommodates the possibility of a weaker convective signal was initially introduced by \cite{spruit_convection_1997} based on the observation that part of the enthalpy flux can be transported by dense downdrafts generated at the surface by abrupt radiative cooling. Due to the Sun's small Prandtl number, such downdrafts could penetrate throughout the whole convective zone \citep{schumacher_colloquium_2020} and, under certain circumstances, produce stably stratified regions.

A scenario in which the heat transport in the solar interior operates under a marginally stable background, i.e., with a subcritical Rayleigh number \citep{chandrasekhar_hydrodynamic_1981}, has been proposed to allow for tilted contours of differential rotation \citep{rempel_solar_2005}. This stable environment was found in the form of the so called Deardorff layers in convection simulations with the radiative conduction governed by Kramers's opacity law \citep{kapyla_extended_2017,brandenburg_stelar_2016,kapyla_simulations_2025}. It has also been evoked to explain the origin and characteristics of the solar polar vortex \citep{bekki_suns_2024,souza-gomes_simulations_2026}, and could also explain the observational absence of large convective structures, \textit{banana cells}, which are recurrent in simulations of rotating convection. This scenario, however, requires a revision of the conditions under which a steady solar-like differential rotation can be sustained in the absence of convection. This task is the primary motivation of the present study.

A convincingly simple theory for explaining the observed differential rotation (DR) profile was proposed by \citet{balbus_differential_2009,balbus_global_2012,balbus_stability_2012,gunderson_model_2019} based on the assumption that the outward convective flows and the observed distribution of angular momentum can only be held steady if they are causally related. Assuming an arbitrary functional relation between the entropy and the rotation rate distributions as well as thermal wind balance (TWB), these models are not only capable of reproducing the observed radial isocontours of rotation, but also indicate that longitudinal motions between the tachocline and the surface are most likely in thermal wind balance. Such TWB hypothesis has also been used to derive the differential rotation of fast rotating stars \citep{rieutord_dynamics_2006}.

More recently, inspired by the resemblance between the solar asphericity and differential rotation profiles, \cite{hester_dynamic_2025} went so far as to propose a balance between the pressure gradients of global shape, the centrifugal forces of global rotation, and the Coriolis forces of the large scale differential motions. Both in this case and the TWB models, the latitudinal distribution of solar winds could be explained without relying on convection in a similar form as the Earth atmosphere winds are explained as a consequence of differential solar heating \citep{marshall_atmosphere_2008}. 

Because an intrinsic baroclinicity may excite horizontal motions, it is still unclear whether the respective thermal wind balance profile could be realized when the rotating flow evolves in the non-linear regime. In this work, we address this question by performing numerical simulations of a spherical shell resembling the bulk of the solar convection zone under the action of a prescribed latitudinal gradient of potential temperature that is, in principle, capable of sustaining the observed solar differential rotation. Here the latitudinal entropy gradient is prescribed as a controlled forcing to test its dynamical consequences rather than to model its physical origin. We explore cases where this layer is either stable or strictly adiabatic, and for completeness, we also study the behavior of convective cases when the environment has such baroclinicity. 

The paper is organized as follows. In section \ref{sec:setup} we describe the analytical formulation of our numerical model as well as the possible TWB solution compatible with the observed solar differential rotation. The results of our simulations are presented in section \ref{sec:results}, and our conclusions and a discussion of the proposed scenario is presented in section \ref{sec:conclusion}.

\section{Numerical Model}
\label{sec:setup}

The numerical experiments conducted in this work consist on the evolution of the solar plasma contained in a spherical shell within the radii $\rmin = 0.72\Rs$ and $\rmax = 0.96\Rs$. We consider a non-magnetic ideal gas to model the plasma's large scale motion. Helioseismic inversions point to a quasi adiabatic stratification in that region, motivating us to consider the following anelastic system \citep{lipps_scale_1982,smolarkiewicz_semi-implicit_2019}:
\begin{align}
	\boldsymbol{\nabla} \cdot \left(\rho_0 \boldsymbol{u}\right)
	 & = 0 \;, \label{eq:anelastic_mass}
	\\
	\frac{D\boldsymbol{u}}{Dt} + 2\boldsymbol{\Omega}\times\boldsymbol{u}
	 & = -\boldsymbol{\nabla}\left(\frac{p'}{\rho_0}\right)+\boldsymbol{g}\frac{\Theta^\prime}{\Theta_0} \; , \label{eq:anelastic_mom}
	\\
	\frac{D\Theta^\prime}{Dt}
	 & = -\boldsymbol{u}\cdot\boldsymbol{\nabla}\Theta_a - \frac{\Theta^\prime - \Theta_{B}}{\tau_{\Theta}}. \label{eq:anelastic_pot}
\end{align}
where $D/Dt=\partial/\partial t + \boldsymbol{u}\cdot\boldsymbol{\nabla}$ is the material derivative and $\boldsymbol{u}=(u,v,w)$ is the velocity field in a rotating frame with $\boldsymbol{\Omega} =\Omega_0(\cos\theta,-\sin\theta, 0)$, where $\Omega_0/2\pi = 445.15$ nHz corresponding to the angular velocity of the radiative zone inferred by helioseismic inversions of \cite{larson_global-mode_2018}. Symbols $p$, $\rho$ and $\Theta$ denote pressure, density, and potential temperature, where the latter is related to the thermodynamic temperature $T$ by $\Theta = T(P_0/P)^{R/c_p}$ with $c_p = 2.5R$ and $P_0$ being the pressure at the bottom of the domain. Primed variables represent perturbations around an ambient state $\{p_a, \rho_a, \Theta_a\}$, whereas variables $\{p_0, \rho_0, \Theta_0 \}$ represent the base state of the anelastic expansion. The variable $\Theta_B$ is a prescribed potential temperature profile described below.

The gravity force, $\bm{g}$, in \cref{eq:anelastic_mom} is purely radial and obeys an inverse square law:
\begin{align}
	\bm{g} = -g_0 \left(\frac{r_{\rm min}}{r}\right)^2\bm{r}
	\label{eq:gravity}
\end{align}
where $\bm{r}$ is the radial unit vector and $g_0 = 511$ ms$^{-2}$ corresponds to the gravitational effect of a spherically symmetric enclosed mass distribution at $r_{min}$.

\begin{figure}
	\includegraphics[width=\linewidth]{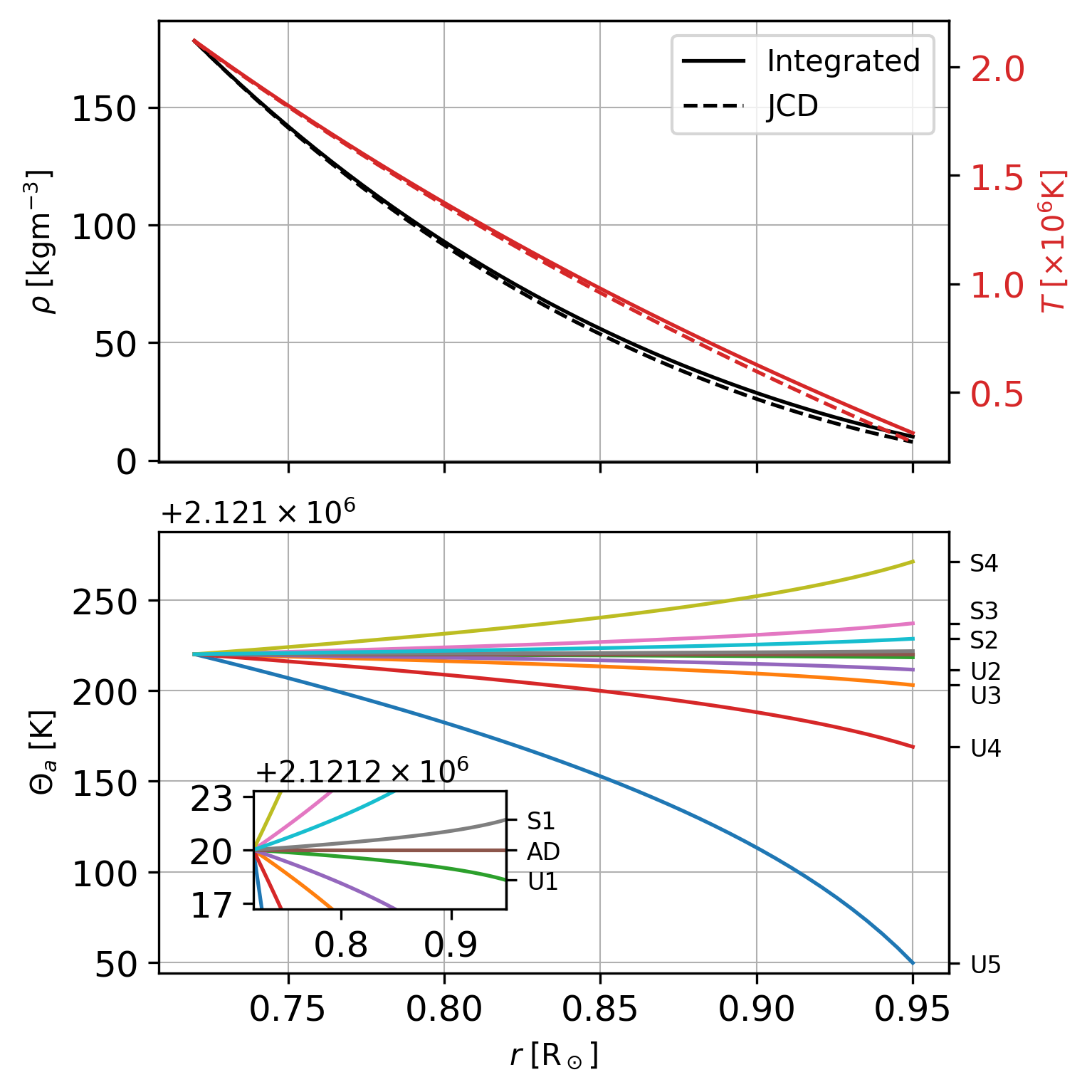}
	\caption{Top panel: polytropic background density, $\rho_0$ (black line) and temperature, $T_0$ (red line), profiles plotted as a solid line.  The corresponding quantities extracted from the solar standard model of \cite{christensen-dalsgaard_current_1996} are presented with dashed lines. Bottom panel: radial profiles of the ambient potential temperature for each class of simulations. The S, U and AD labels refer, respectively, to stable, unstable and adiabatic stratifications, and the corresponding numbers are mapped to the same classification detailed in \cref{tb:simulations}. The inset on the bottom figure is a zoom to a narrower range in the y-axis.}
	\label{fig:strat}
\end{figure}

We assume the background temperature $T$ and density $\rho$ are spherically symmetric and given by the numerical integration of polytropic relation in hydrostatic balance,
\begin{align}
	\frac{d T_i}{d r}   & = - \frac{g}{R_g\left({m}_i+1\right)},
	\label{eq:poly_temp}
	\\
	\frac{d\rho_i}{d r} & = -\frac{\rho_i}{T_i}\left(\frac{g}{R_g}+\frac{d T_i}{d r}\right),
	\label{eq:poly_rho}
\end{align}
where $R_g = 13732.0$ is the gas constant for a fully ionized hydrogen mixture and the index $i = 0, a$ denotes the base and ambient states respectively. For our case of a perfect gas, the polytropic index, $m_i$, controls the stratification, resulting in Schwarzschild stable (unstable) states for values $m > 1.5$ ($m < 1.5$). In all cases, the base state polytropic index is fixed to $m_0 = 1.5$ which corresponds to an adiabatic stratification, represented in the first panel of \cref{fig:strat}. Each profile is integrated radially outward using reference values at $r = r_{min}$ from the solar standard model \citep{christensen-dalsgaard_current_1996}. 

Notice the inclusion of a Newtonian cooling term in the r.h.s. of \cref{eq:anelastic_pot} whose role is to nudge the evolution of the model towards a prescribed profile $\Theta \rightarrow \Theta_a(r) + \Theta_B(\theta, r)$ on a timescale of $\tau$. This allows us to study the evolution of the prognostic variables under the constraints of a prescribed radial thermal stratification and baroclinicity. Such technique is a common practice in global simulations \citep{cossette_supergranulation_2016,cossette_magnetically_2017,nogueira_numerical_2022,guerrero_implicit_2022} and is generally known as \textit{nudging operation} in the data assimilation literature \citep{lakshmivarahan_nudging_2013}. Because the relaxation timescale (2.4–4.9 years) is much longer than the dynamical timescales, the nudging acts as a weak thermodynamic constraint; it does not directly prescribe the flow evolution. Its physical interpretation is that of a \textit{wavelength agnostic} diffusivity, since in Fourier space it takes the form of $-\tau^{-1} \hat{\Theta}^\prime$ as opposed to the standard Fickian diffusion, $-\nu k^2 \hat{\Theta}^\prime$.

The nudging target profile $\Theta_B$ implements an axissymmetric potential temperature distribution in thermal wind balance with a velocity profile consisting only of differential rotation $\bm{u} = (u_0, 0, 0)$, obtained by fitting the helioseismic observations \citep{larson_global-mode_2018} into $u_0 = a_0(r) + a_2(r)\cos^2(\theta) + a_4(r)\cos^4(\theta)$. Both $\Omega = u_0/(r\sin(\theta))$ and its corresponding profile of $\Theta_B$ are presented respectively in the left and right panels of Fig.\ref{fig:heating}. As detailed in Appendix \ref{apd:heating_profile}, the construction of such state is defined up to an arbitrary radial term, which we fix by imposing a vanishing averaged stratification $\int \Theta_B(\theta,r) r d \theta = 0$, allowing the ambient state $\Theta_a$ to control the forced thermal stratification.

\begin{figure}
	\includegraphics[width=\linewidth]{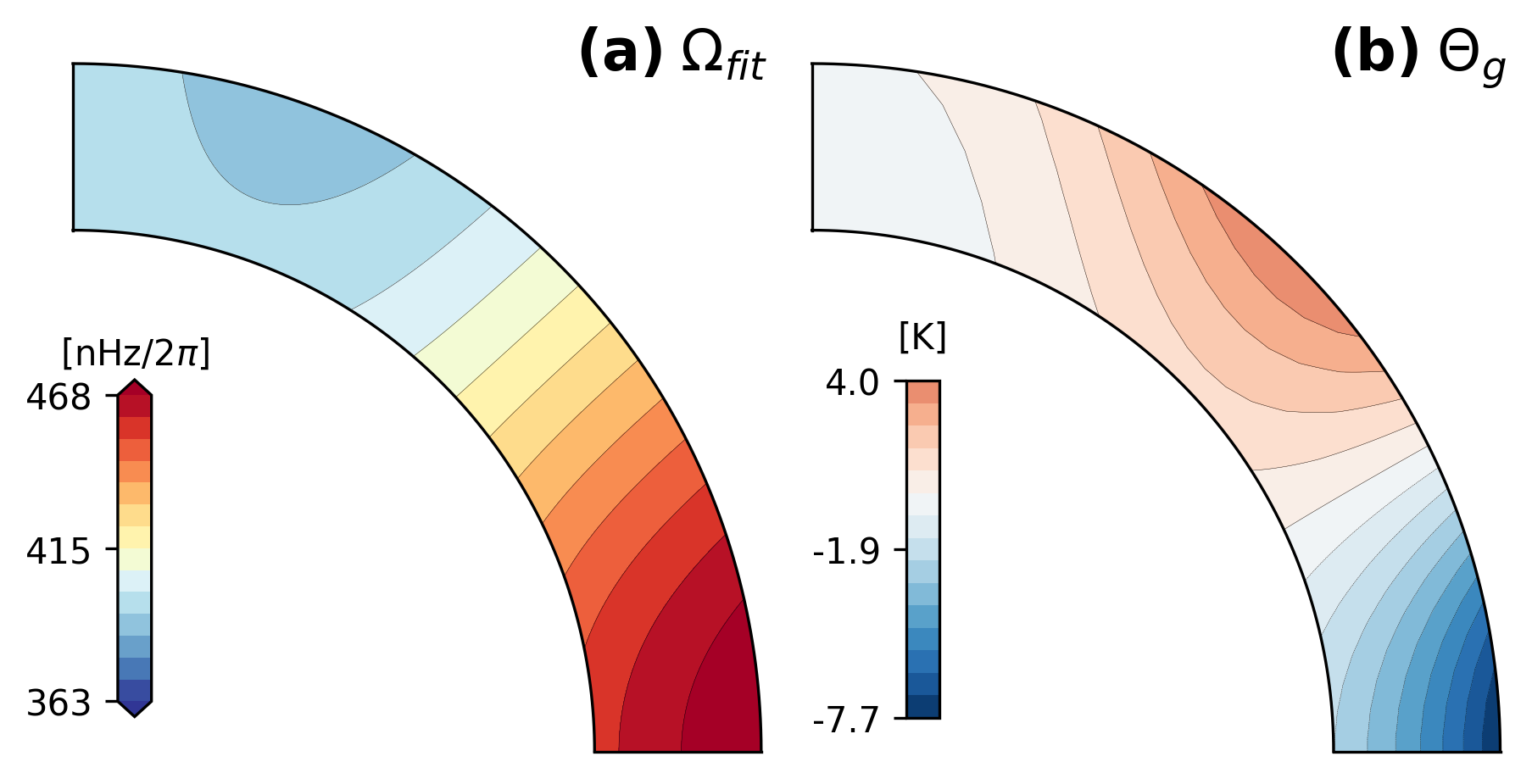}
	\caption{Angular velocities fitted from the heliosseismic data \citep{larson_global-mode_2018} into \cref{eq:obs_fit_form,eq:obs_fit_constrain} (left), color limits are fixed to match the heliosseismic data. Potential temperature profile obtained after integration of \cref{eq:heating}, using the fitted velocities, and imposition of a vanishing latitudinal average (right).}
	\label{fig:heating}
\end{figure}

Boundary conditions in both inner and outer shells are impermeable and stress-free. The potential temperature fluctuations are forced to vanish at both boundaries, all together that implies:
\begin{align*}
	w(\phi,\theta,\rmin)               & = w(\phi,\theta,\rmax) = 0                \\
	\Theta^{\prime}(\phi,\theta,\rmin) & = \Theta^{\prime}(\phi,\theta,\rmax)  = 0 \;.
\end{align*}

The governing equations \labelcref{eq:anelastic_mass,eq:anelastic_mom,eq:anelastic_pot} are solved in a spherical grid, $(n_{\phi}, n_{\theta}, n_r)= (128,64,48)$, using the EULAG-MHD code \citep{prusa_eulag_2008,smolarkiewicz_eulag_2013,smolarkiewicz_semi-implicit_2019}, which employs an anti-diffusive, non-oscilatory and semi-implicit time integration scheme as well as a preconditioned GCR(k) algorithm for solving the elliptic pressure equation.

\section{Results}
\label{sec:results}

Motivated by observational evidence of weak convective motions and by the possibility that the convection zone contains subadiabatic layers, we aim to verify whether a thermal baroclinic forcing may lead to the development of a steady differential rotation. To this end, we perform a series of simulations in which we vary the superadiabaticity of the layer, $\Delta\nabla = -{(d\ln \Theta)}/{(d\ln P)}$, through the polytropic index, $m_a$, in \cref{eq:poly_rho,eq:poly_temp}, as well as the forcing timescale, $\tau$, in \cref{eq:anelastic_pot}. The different choices of $m_a$, listed in \cref{tb:simulations}, correspond to convectively unstable, neutral, or stable stratifications. We note that some unstable simulations require a smaller time step due to the development of large vertical velocities. The corresponding integrated ambient profile $\Theta_a$ for all simulations performed in this study is shown in the bottom panel of \cref{fig:strat}.

\begin{deluxetable*}{lccc}
    \digitalasset
    \tablewidth{0pt}
    \tablecaption{Labels and respective parameters of all simulations performed. The naming convention follows the stability of the resulting $\Theta_a^{s}(r)$ profile given the polytropic index, $m$. The superadiabaticity is computed as a radial average of $\Delta\nabla= -{d(\ln\Theta)}/{d(\ln P)}$. For each polytropic index we performed simulations with two different values of $\tau_{\Theta}$, $7.8\times10^7$ ($\approx 2.4$ years), $1.5\times10^8$ ($\approx 4.9$ years). The last column shows the time step used in each group of simulations, with $\delta t_0=3000$ s.
	\label{tb:simulations}}
     
    \tablehead{
    \colhead{Label} & 
    \colhead{Polytropic index $m$} & 
    \colhead{Ambient superabadicity $\Delta\nabla$} &
    \colhead{Time step}
    }
    \startdata
    U5 & $1.5 - 1 \times 10^{-4}$ & $ 1.6 \times 10^{-5}$  & $\delta t_0/3$ \\
    U4 & $1.5 - 3 \times 10^{-5}$ & $ 4.8 \times 10^{-6}$  & $\delta t_0/2$ \\
    U3 & $1.5 - 1 \times 10^{-5}$ & $ 1.6 \times 10^{-6}$  & $\delta t_0/2$ \\
    U2 & $1.5 - 5 \times 10^{-6}$ & $ 8.0 \times 10^{-7}$  & $\delta t_0/2$ \\
    U1 & $1.5 - 1 \times 10^{-6}$ & $ 1.5 \times 10^{-7}$  & $\delta t_0$   \\
    AD & $1.5$ & $ 0.0$                   & $\delta t_0$ \\
    S1 & $1.5 + 1 \times 10^{-6}$ & $ -1.5 \times 10^{-7}$ & $\delta t_0$ \\
    S2 & $1.5 + 5 \times 10^{-6}$ & $ -8.0 \times 10^{-7}$ & $\delta t_0$ \\
    S3 & $1.5 + 1 \times 10^{-5}$ & $ -8.0 \times 10^{-6}$ & $\delta t_0$ \\
    S4 & $1.5 + 3 \times 10^{-5}$ & $ -1.5 \times 10^{-6}$ & $\delta t_0$ \\
    \enddata
\end{deluxetable*}

For each value of $m_{a}$, we perform two simulations with $\tau = 2.4$ and $\tau = 4.9$ years (see \cref{tb:simulations}). This yields a total of 20 simulations, each initialized with the fluid at rest, $\bm{u} = 0$, and a random uniform $\Theta^{\prime}$ perturbation ranging from $-0.1$ K to $0.1$ K. Such perturbation is sufficient to excite the development of large scale flows which are evolved for $\approx 456$ years. At the first years of integration, spatial averages of $\Theta'$ and $|u|$ increase linearly before attaining a steady plateau. The temporal averages presented below are computed over the final 76 years.

\begin{figure}
	\includegraphics[width=\columnwidth]{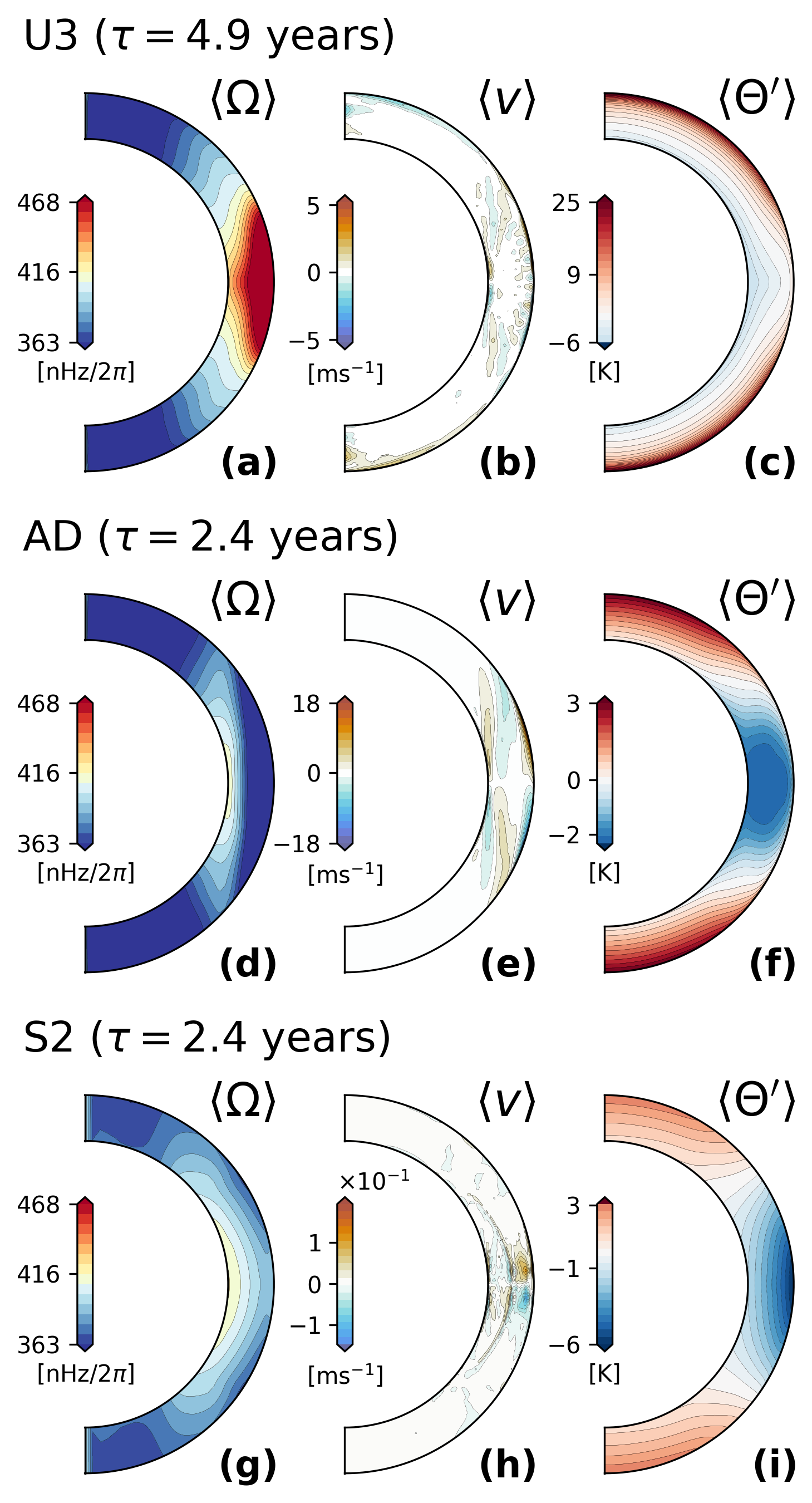}
	\caption{Time and longitudinal averages of the rotation rate 
    (a, d, g), latitudinal velocity (b, e, h) and potential temperature perturbation (c, f, i) fields for three characteristic simulations chosen as so to represent the total set of simulations. Color levels on the first column is adjusted to match the limits of heliosseismic data \citep{larson_global-mode_2018}. Each row corresponds to a different simulation named after the conventions of \cref{tb:simulations}. }
	\label{fig:selected}
\end{figure}

To facilitate the discussion of our results, we identified a subset of three simulations whose final, time-averaged, differential rotation profiles are morphologically representative of all other simulations. This subset contains simulations with unstable, adiabatic and subadiabatic domains. In \cref{fig:selected}, we show time and longitudinal averages of the differential rotation, $\left<\Omega\right>$ (left panels), the latitudinal velocity, $\left<v\right>$ (middle panels), and the potential temperature perturbation $\left<\Theta'\right>$ (right panels), for the three simulations of this subset. An equivalent plot for all simulations of \cref{tb:simulations} is provided in Appendix \ref{apd:all_sims}.

Regardless of the superadiabaticity of the domain, $\Delta\nabla$, all simulations exhibit an overall trend of $\left<\Omega\right>$ to increase from pole to equator and have radially aligned contours at mid latitudes, defining a solar-like character. This is congruent with the tendency of the forced baroclinicity to excite equator-wards flows. On the other hand, the radial behavior of $\left< \Omega \right>$ is sensitive to particular choice of $(\Delta\nabla, \tau$) and is strongly correlated with the meridional flow's amplitude and structure.

Weakly convective models, such as the one represented in the first row of \cref{fig:selected}, are the only ones capable of reproducing the observed radial increase of $\left<\Omega\right>$ at low latitudes. They are also the only models that display a substantial change in velocity fields averages when the stratification's stiffness, controlled by $\tau$, is altered. The solar-like differential rotation in these models is accompanied with weak and sparse meridional cells, a pattern previously reported in literature \citep{featherstone_meridional_2015}. While they work to speed up the equator, the Reynolds stresses are responsible for torques capable of  maintaining the radial increase of the rotation rate (see next section). This subtle balance is lost when the superabadicity is increased (see the U5 models in \cref{fig:all}) suggesting that convection, if present in the sun, has to be weak in order to not diffuse away those structures.

As the model's superabadicity approaches the adiabatic value ($\Delta\nabla \approx 0$), the influence of stratification on meridional motion becomes less significant. This allows for the appearance of strong meridional cells aligned with the rotation axis, which tend to confine the fast rotating regions inside a tangent cylinder. As shown in \S \ref{sec:twb}, those motions originate from a gyroscopic pumping effect due an excess of baroclinicity. This trend persists up to the stable model S2, displayed in the last row of \cref{fig:selected}.

A further increase of the stratification's stability from the models S2 to S4 inhibits the radial movement of potential temperature fluctuations. This explains the significant reduction of the meridional flow's amplitude and its tendency to organize in a stack of thin cells. As a consequence, the net radial torques become smaller as so the overall spatial rotation contrast.

Regarding the averaged potential temperature perturbation $\left<\Theta'\right>$, all simulations display a positive latitudinal gradient as induced by the forcing process. Unstable simulations exhibit a positive radial gradient, due to a tendency of convection to flatten the total entropy gradient $\Theta_a + \Theta'$.

\subsection{Integrated torques}
\label{sec:reynolds}

The role of the meridional flow in the establishment of a steady differential rotation is
better described by decomposing the fluid velocity components into average and fluctuating components $\bm{u} = \bm{\overline{u}} + \bm{u'}$, where the bar denotes a longitudinal average. The decomposed fields are substituted in the equation of evolution for the mean angular momentum $\bm{\mathcal{L}} = \lambda \rho \bm{\overline{u}}$ deduced from \cref{eq:anelastic_mom}, yielding \citep{brun_turbulent_2002,guerrero_implicit_2022}:
\begin{align}
	\partial_t \mathcal{L} = -\nabla \cdot (F_r^{RS} + F_r^{MC} + F_\theta^{RS} + F_\theta^{MC}) \;,
	\label{eq:l_evolution}
\end{align}
where the terms under the divergence are explicitly written in terms of the longitudinal, latitudinal and radial velocity components $u, v$ and $w$ as,
\begin{equation}
	\begin{split}
		F_r^{MC}      & = \rho_0\lambda (\bm{\overline u} + \lambda\Omega_0)\overline{w} \;, \\
		F_r^{RS}      & = \rho_0\lambda \overline{u'w'}                                \;,  \\
		F_\theta^{MC} & = \rho_0\lambda (\bm{\overline u} + \lambda\Omega_0)\overline{v} \;, \\
		F_\theta^{RS} & = \rho_0\lambda \overline{u'v'} \;.	\end{split}
	\label{eq:rs_def}
\end{equation}

These terms represent the radial and latitudinal advection of angular momentum, denoted respectively by the $r$ and $\theta$ subscripts. The meridional circulation terms, denoted by the MC superscript, encompass the contribution of the large scale component of the flow ($m = 0$ in a spherical harmonics decomposition) to the transport of angular momentum, whereas the Reynolds stress, denoted by the RS superscript, sums the contribution of correlating fluctuations ($m > 0$) around the longitudinal average.

The role the angular momentum fluxes, defined in \cref{eq:rs_def}, is easily understood through the integrated torques:
\begin{align}
	\begin{split}
		I_r(r)           & = \int_0^\pi F_r(r, \theta) r^2\sin\theta d\theta \;,
		\\
		I_\theta(\theta) & = \int_0^\pi F_\theta(r, \theta) r\sin\theta d r \;.
	\end{split}
	\label{eq:rs_int}
\end{align}
where $I_r$ and $I_\theta$ denote the net torque across spherical and conical surfaces, respectively. Since boundaries are stress-free, the total angular momentum is constant and may only be redistributed by the action of internal torques.

\begin{figure*}
	\includegraphics[width=\textwidth]{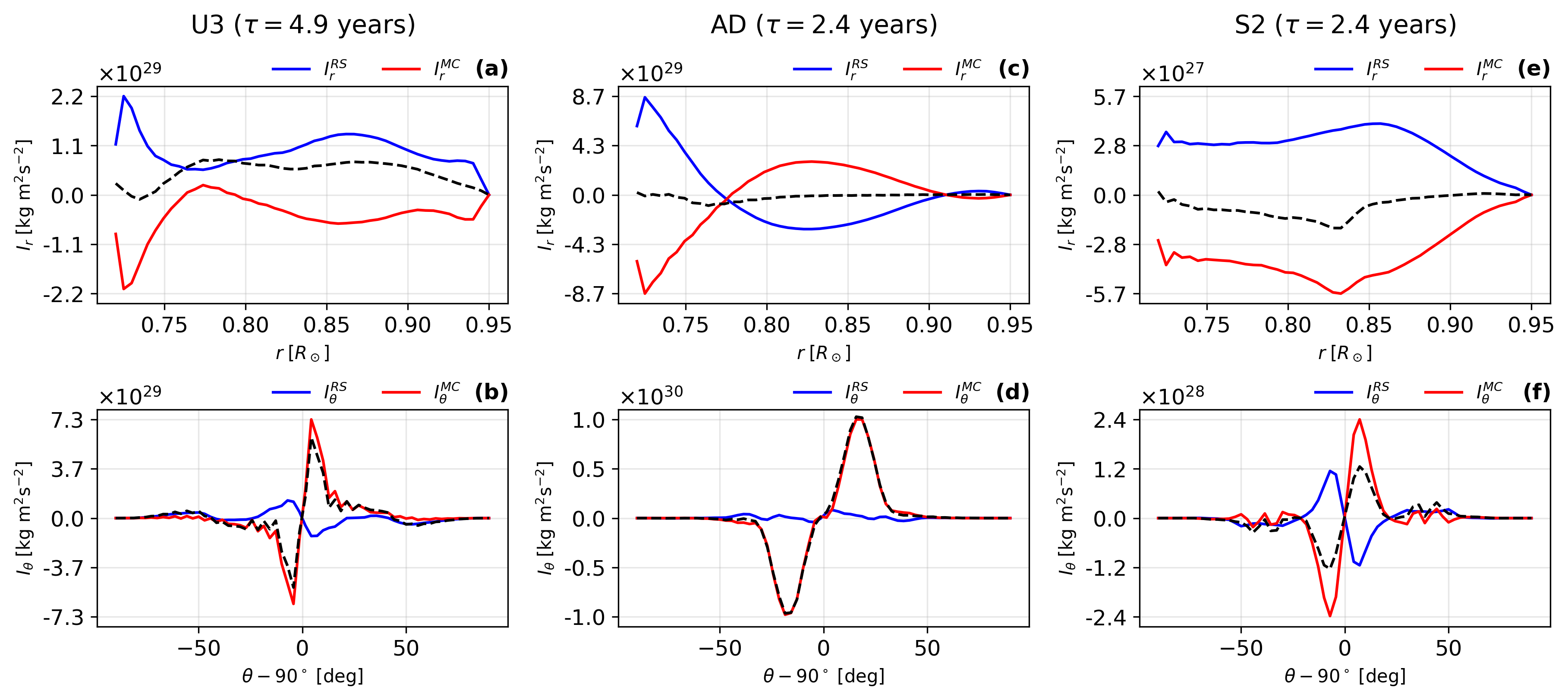}
	\caption{Time averaged radial (top row) and latitudinal (bottom row) integrals of the Reynolds stresses defined in \cref{eq:rs_int} for the same simulations of \cref{fig:selected}.}
	\label{fig:selected_reynolds}
\end{figure*}

The radial and latitudinal profiles of those integrated torques for the same set of simulations presented in \cref{fig:selected} are respectively plotted in the upper and lower panels of \cref{fig:selected_reynolds}. In the three simulations, the meridional circulation contribution to $I_{\theta}$, indicated by red lines, is positive (negative) in the northern (southern) hemisphere. In effect, this accelerates the equator, as opposed to the Reynolds stresses that have opposite sign or approximately vanish as in the case AD. Among these simulations, the stable case, S2, is the one with the smallest torques, due to its stratification resistance against buoyant radial motion.

A residual torque, plotted as a dashed line in \cref{fig:selected_reynolds}, indicates a tight but not strict balance 
in the radial direction. Keeping the attention on this component, one notes that the angular momentum flux peeks close to the bottom of the domain in unstable and adiabatic simulations whereas it stays approximately independent of the radius in the stable case. Inspection of the longitudinal averages of $-\nabla \cdot (\mathcal{F}_{r}^{RS} + \mathcal{F}_{\theta}^{RS})$ and $-\nabla \cdot (\mathcal{F}_{r}^{MC} + \mathcal{F}_{\theta}^{MC})$ (not shown) reveals that such peaks occur predominantly at equatorial regions. Although not shown here, such trend is shared among simulations with the same stability group.

\subsection{Thermal wind balance}
\label{sec:twb}

As shown in the previous sections, the steady differential rotation profile achieved in the simulations departs significantly from the one compatible with the imposed baroclinicity. By evaluating the net torques, we concluded that this is a result of the action of developed meridional circulation cells. The origin of such flow can be enlighten by looking at the longitudinally averaged equation for $\left(\nabla \times \bm{v}\right)_\phi$ \citep{kitchatinov_theory_2012,choudhuri_meridional_2020}:
\begin{align}
\mathcal{M} &=
r \sin(\theta) \frac{\partial \Omega^2}{\partial z} - \frac{g}{\Theta_0 r }\frac{\partial \Theta'}{\partial\theta} 
 = \mathcal{C} - \mathcal{B} 
\label{eq:twb} 
\end{align} 
where $\mathcal{M}$ encompasses the contributions from the meridional circulation and viscous effects. It follows from \cref{eq:twb} that, assuming a steady perfect fluid, an imbalance on the thermal wind relation ($\mathcal{C} \neq \mathcal{B}$) must be compensated by a meridional flow ($\mathcal{M} \neq 0$), a mechanism known as gyroscopic pumping \citep{garaud_gyroscopic_2010,miesch_gyroscopic_2011}.

\begin{figure}
	\includegraphics[width=\columnwidth]{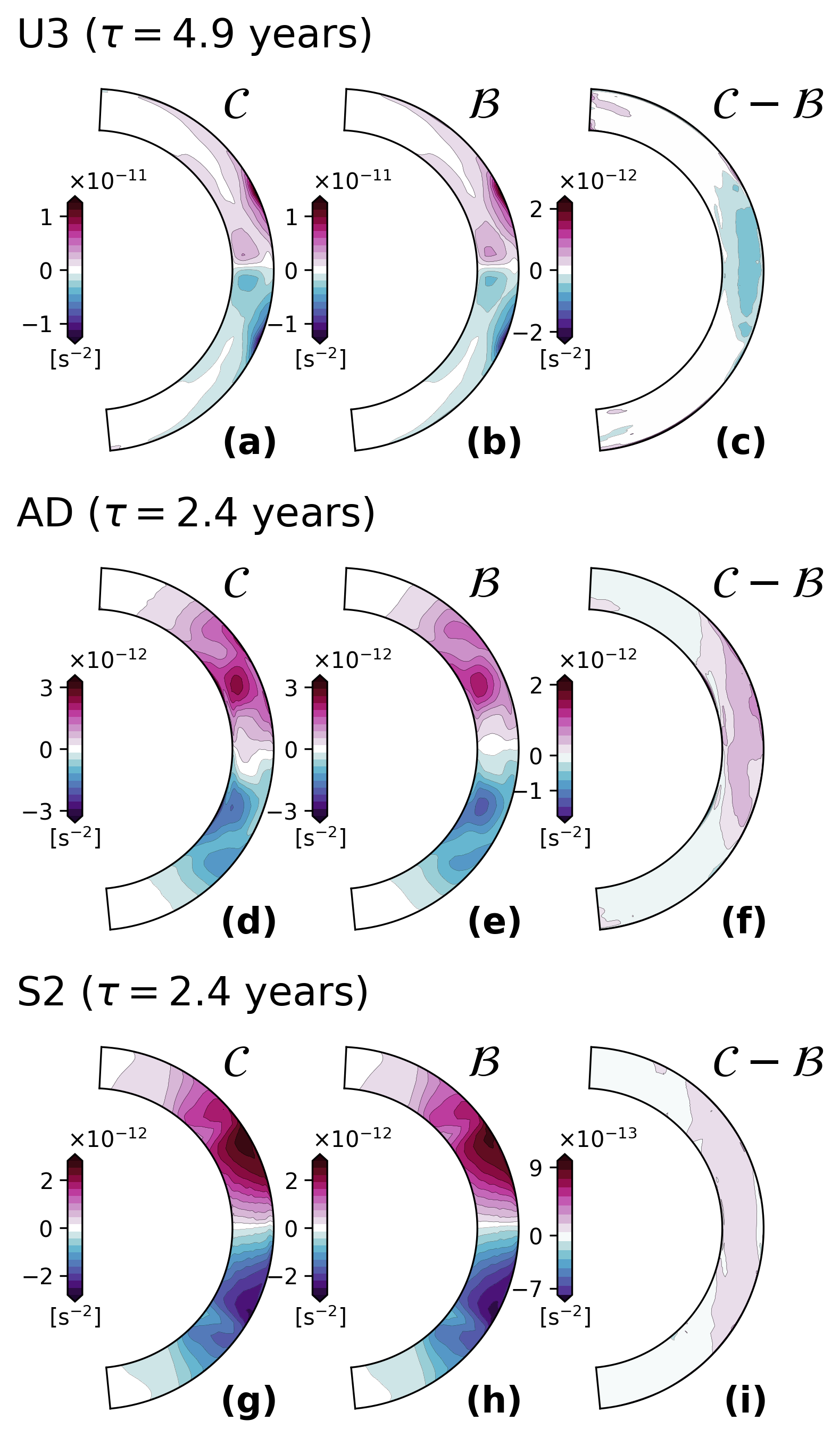}
        \caption{Time and longitudinal averages of the centrifugal (panels (a), (d), (g)) and baroclinic (panels (b), (e), (h)) terms, defined in \cref{eq:twb}, as well as their difference (panels (c), (f), (i)) for the same set simulations of \cref{fig:selected}.}
	\label{fig:twb}
\end{figure}

In \cref{fig:twb} we present the longitudinal and time averages of $\mathcal{C}$ (left panels) and $\mathcal{B}$ (middle panels) as well as their difference (right panels). The residual of the two terms is one order of magnitude smaller than $\mathcal{C}$ and $\mathcal{B}$ and appears at equatorial latitudes. Such small, yet significant, imbalance explains the appearance of meridional cells, which had otherwise been neglected in the construction of the forced baroclinicity. 

Is is worth noting that, in a steady state, such cells must balance the angular momentum transport by turbulent stresses \citep{miesch_gyroscopic_2011}. Such requirement is clearly satisfied by the meridional circulation's torques shown by the red lines in \cref{fig:selected_reynolds} as well as the cell's orientation indicated in the middle panels of \cref{fig:selected}, endorsing that they originate from a gyroscopic pumped effect.

\section{Concluding remarks}
\label{sec:conclusion}

In this work we investigated whether the observed solar differential rotation can be sustained, without requiring dominant Reynolds stresses associated with vigorous turbulent convection, but by imposing a large-scale latitudinal entropy gradient (externally prescribed) consistent with thermal wind balance. Using global anelastic simulations spanning stable, adiabatic, and superadiabatic stratifications within the bulk of the convection zone ($0.72–0.96 R_\odot$), we demonstrate that solar-like differential rotation, characterized by a fast equator and slower poles, can be generated and maintained without relying on Reynolds stresses from strong turbulent convection.

In particular, weakly (large $\tau$) and stably ($\Delta\nabla < 0$) convective cases reproduce the observed fast-equator rotation profile throughout most of the convection zone as well as the mid-latitude radially aligned contours. Strongly superadiabatic cases, by contrast, tend toward cylindrical (Taylor-Proudman) rotation contours, indicating that vigorous convection acts to reduce the imposed baroclinicity. These results suggest that large-scale baroclinic forcing, rather than turbulent angular momentum transport alone, may be the primary constraint shaping the solar rotation profile in the convection zone interior. However, determining how this large-scale entropy anisotropy is established in the solar interior remains an important topic for future investigation.

Our findings are consistent with thermal wind interpretations of the helioseismic rotation structure \citep{balbus_simple_2009,balbus_global_2012, balbus_stability_2012,gunderson_model_2019}. They are similar to the recent dynamic equilibrium model of \citep{hester_dynamic_2025}, which also did not require convection. However, whereas the latter relied on a global pressure gradient asphericity, we used a global scale entropy gradient as the primary forcing. While the physical origin of the required entropy anisotropy here remains unresolved, several mechanisms proposed in the literature remain viable, including rotational modulation of turbulent heat transport \citep{kitchatinov_differential_1995}, tachocline-convection zone coupling \citep{rempel_solar_2005}, and entropy rain driven by surface cooling \citep{spruit_convection_1997,brandenburg_stelar_2016}. The present work does not prescribe the source of this anisotropy; rather, it establishes its dynamical viability as a sustaining mechanism for differential rotation.

These results bear directly on the so-called convective conundrum \citep{hanasoge_anomalously_2012,omara_velocity_2016,birch_solar_2024,stefan_time-dependence_2026}, wherein helioseismic constraints imply convective amplitudes significantly smaller than those predicted by mixing-length theory. By demonstrating that solar-like differential rotation does not require strong Reynolds stresses, this study demonstrates a physically viable pathway toward reconciling the observed rotation profile with reduced convective intensity and marginal stability in the convection zone interior.

More broadly, our results suggest that solar differential rotation may be fundamentally governed by global thermodynamic balance, with convection playing a secondary or modulating role. If so, similar baroclinic constraints may operate in other solar-type stars. Extending this framework to magnetohydrodynamic regimes and assessing its implications for dynamo action constitute important next steps.

In summary, the solar convection zone may be closer to marginal stability than traditionally assumed, with differential rotation emerging as a manifestation of large-scale baroclinic equilibrium. This perspective provides a viable alternative to convection-dominated paradigms and contributes toward resolving long-standing discrepancies between observations, theory, and numerical simulations.
a

\begin{acknowledgments}
L.S.M. and M.D. acknowledges partial support from NASA sub-award from Stanford's NASA-DRIVE Center award 80NSSC22M0162.. G.G. acknowledges financial support from CNPq grant 309695/2025-2. L.S.M. acknowledge financial support from CNPq.
This work is supported by the NSF National Center for Atmospheric Research, which is a major
facility sponsored by the National Science Foundation (NSF) under cooperative agreement 1852977. Computational resources supporting this work were provided by the NASA High-End Computing Program through the NASA Advanced Supercomputing Division at Ames Research Center and the Computational and NCAR's Information Systems Laboratory \citep{computational_and_information_systems_laboratory_derecho_2023}.
\end{acknowledgments}

\appendix

\section{Thermally induced differential rotation}
\label{apd:heating_profile}

To obtain an entropy distribution compatible with the differential rotation profile inferred by helioseismic measurements, we integrate the thermal wind balance relation given by \cref{eq:twb} when $\mathcal{B} = \mathcal{C}$ assuming spherically symmetric $g$ and $\Theta_0$: 
\begin{align}
    \Theta_B = C(r) - \frac{\Theta_0}{g} \int{\frac{\partial\Omega^2}{\partial z} rd\theta}
    \label{eq:heating}
\end{align}
where $C(r)$ is an arbitrary radial function. Such integrated profile also represents the potential temperature distribution required to sustains a velocity field $(u_o, 0, 0)$ in geostrophic balance. 

To evaluate $\Theta_B$, we first interpolate the observed differential rotation $u_0(r, \theta)$, inferred by helioseismology, into a functional form of the type:
\begin{align}
	\begin{split}
		u(r, \theta) = & (\alpha_0 r + \beta_0)
		+ (\alpha_2r + \beta_2) \cos^2(\theta) + (\alpha_4 r + \beta_4) \cos^4\theta
	\end{split}
	\label{eq:obs_fit_form}
\end{align}
The usage of a smooth velocity field given by \cref{eq:obs_fit_form} implies in a smooth ambient profile, which simplifies the analysis of the final steady states $\Theta$. We further reduce the number of ajusted coefficients in \cref{eq:obs_fit_form} by imposing that $u(r, \theta) = 0$ at each pole. This is translated mathematically to:
\begin{equation}
	\begin{aligned}
		\alpha_4 & = -(\alpha_2 + \alpha_0) \\
		\beta_4  & = -(\beta_2 + \beta_0)
	\end{aligned}
	\label{eq:obs_fit_constrain}
\end{equation}
Substituting the relations of \cref{eq:obs_fit_constrain} back to \cref{eq:obs_fit_form}, the resulting expression can be fitted using a multiple linear regression where each independent variable is a linear function of one $\cos^2(\theta)$ or $\cos^4(\theta)$. The corresponding angular velocity is shown in the left panel of \cref{fig:heating}. Finally, the $\Theta_B$ may be obtained by straight integration of \cref{eq:heating} up to a radial constant.

\section{All simulations}
\label{apd:all_sims}

In \cref{fig:all} we display a time and longitudinal averages of the same fields as \cref{fig:selected} but for all simulations listed at \cref{tb:simulations}.

\begin{figure}[ht]
 \centering
  \begin{subfigure}[b]{0.9\textwidth}
    \includegraphics[width=\textwidth]{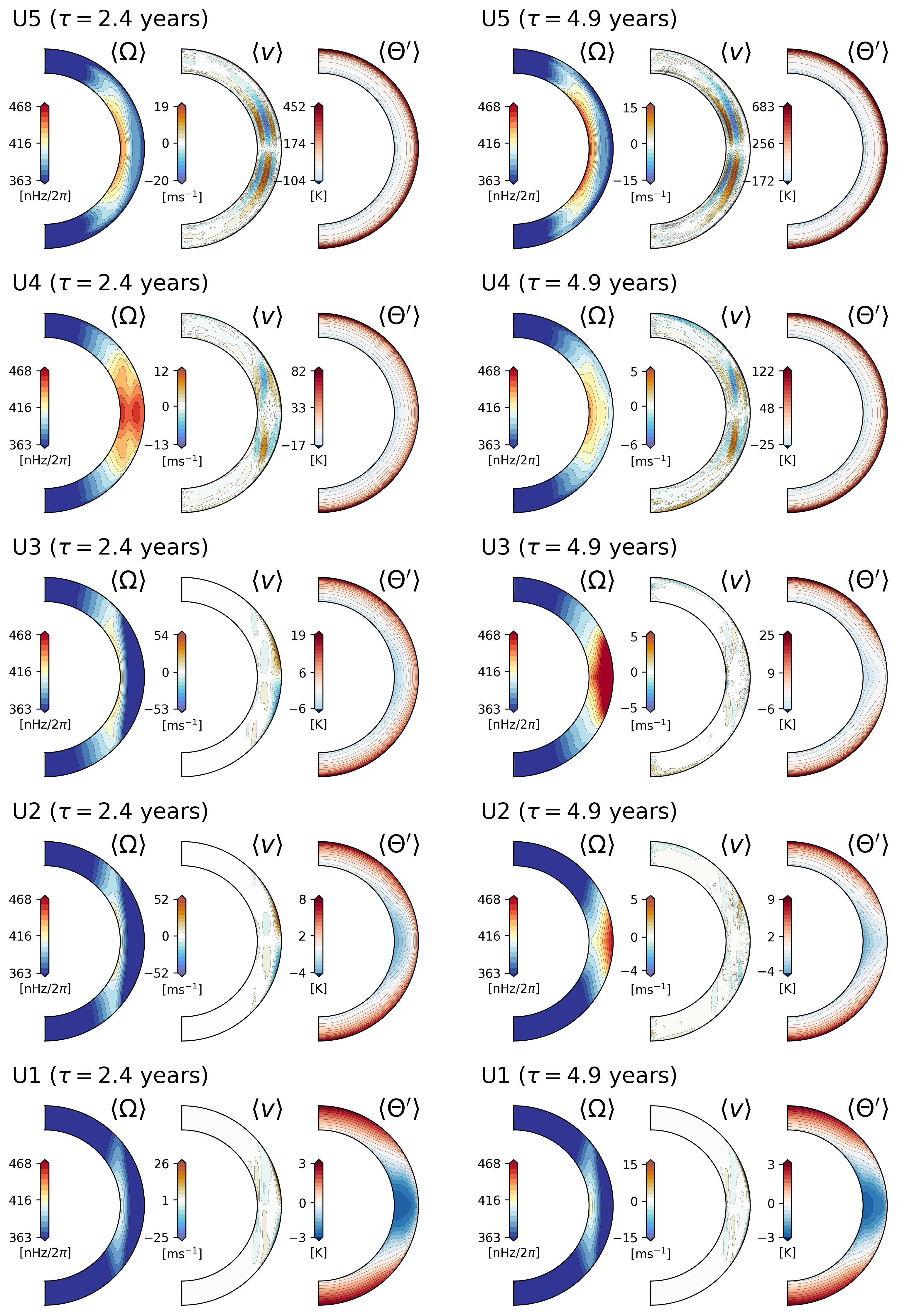}
  \end{subfigure}
\end{figure}

\begin{figure}[ht]
    \ContinuedFloat
    \centering
    \begin{subfigure}[b]{0.9\textwidth}
        \includegraphics[width=\textwidth]{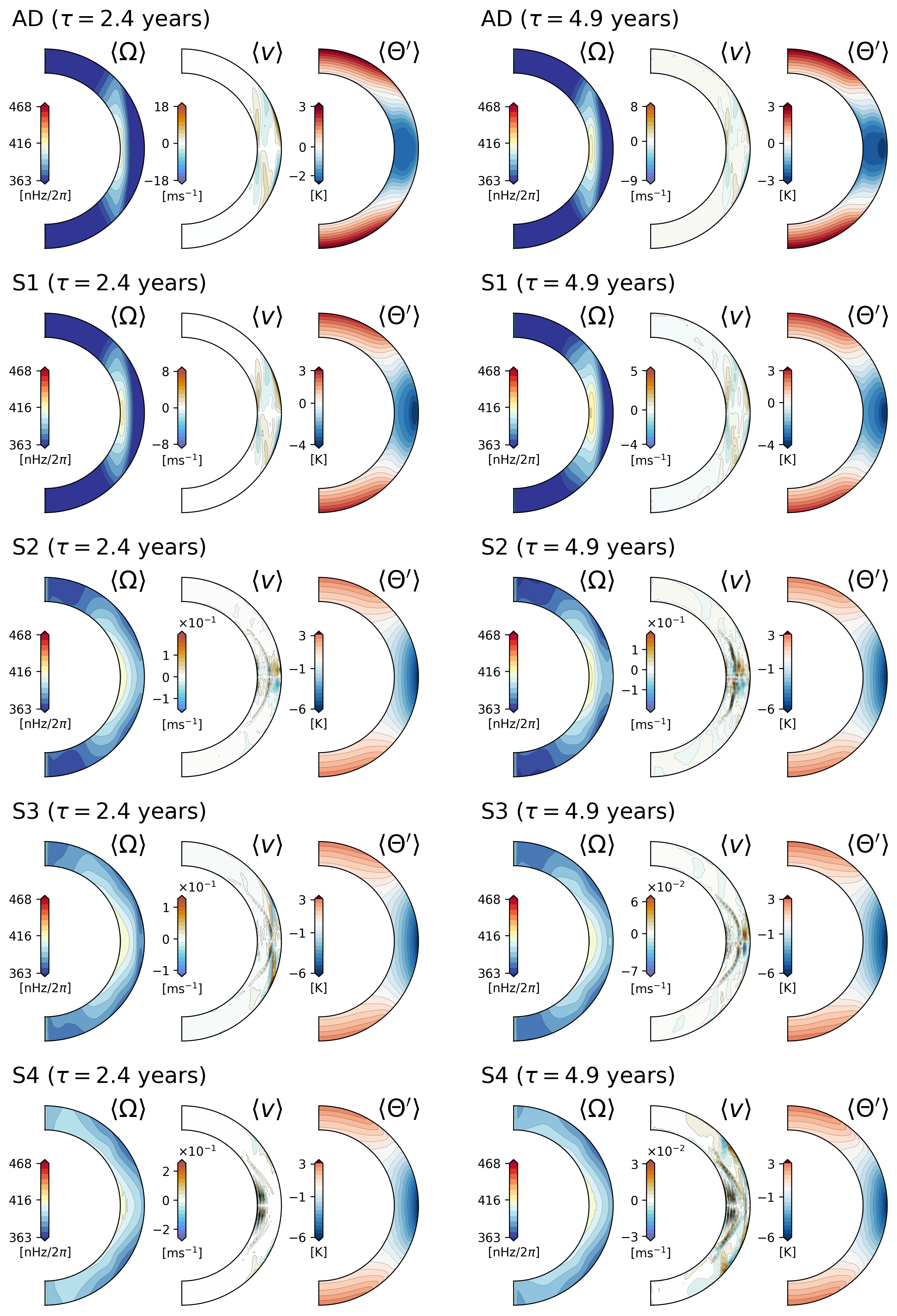}
    \end{subfigure}
    \caption{Time and longitudinal averages of the rotation rate, latitudinal velocity and potential temperature perturbation for the models listed in \cref{tb:simulations} with two forcing timescales $\tau$.}
  \label{fig:all}
\end{figure}

\bibliography{references}{}
\bibliographystyle{aasjournalv7}

\end{document}